\documentclass[a4paper,12pt]{article}
\usepackage{amsmath,indentfirst,amsfonts}
\usepackage[left=2cm,right=2cm, top=2cm,bottom=2cm,bindingoffset=0cm]{geometry}
\usepackage[english]{babel}
\begin{document}
\begin{titlepage}

\begin{tabular}{r}
ITEP - 50/13
\end{tabular}
\vskip1cm
\begin{tabular}{r}
to appear in Theor. Math. Phys.
\end{tabular}
\vskip2cm
\centerline{\large \bf Semiclassical treatment of pair creation in de Sitter space}

\vspace{3cm}

\centerline{ Sergey Guts$^a$}

\vspace{20mm}

\centerline{\it $^a$Institute of Theoretical and Experimental Physics }
\centerline{\it  B. Cheremushkinskaya ul. 25, 117259 Moscow, Russia}

\def\thefootnote{\fnsymbol{footnote}}%
\vspace{2cm}

\centerline{\bf Abstract}

\vspace{5mm}

We study a massive scalar field theory in de Sitter space. Using worldline instanton approach, we calculate probability of pair production in weak-field limit. In addition to exponential factor we derive pre-exponential factor. Within this approach the vanishing probability for odd-dimensional de Sitter space gets a clear geometrical interpretation. We find leading contribution to imaginary part of two point correlator in $\phi^3$-theory.

\end{titlepage}

 \section{Introduction}
 \label{intro}
 Semiclassical methods are a powerful tool to analyze non-perturbative phenomena  in modern quantum field theory. A  false vacuum decay is one of the most well-known examples \cite{colemanClassic},\cite{colemanCorrections}, \cite{Kobzarev:1974cp},\cite{Selivanov:1985vt}. Equations of motion produce a solution, which interpolates between true and false vacua and probability of false vacuum decay is proportional to exponent of the Euclidean action evaluated on this solution.

 Similar methods can be applied when we study particle production in external background. The most well-studied case is Schwinger effect  - a spontaneous electron-positron pair creation in electric field \cite{Schwinger},\cite{affleck},\cite{Dunne:2006st}. The probability of pair production can be expressed in terms of one-loop effective action. The solutions to saddle point equation, called \textit{worldline instantons}, can be interpreted as trajectories of particle moving in Euclidean space.

Strong exponential suppression of such processes became a reason for induced processes research. For a number of cases such processes turned out to be much more probable \cite{Bulycheva:2011gs},\cite{Monin:2009aj},\cite{Gorsky:2005yq}.

 We will be interested in a similar process of pair creation in external gravitational field. Recently this problem has been studied both from quasiclassical \cite{Morozov} and kinetic equation viewpoints \cite{KrotovPolyakov},\cite{Akhmedov:2013xka}. In this article we will explore a model example of self-interacting scalar field theory in de Sitter space. Using the Euclidean action and worldline instanton, we analyze the rate of spontaneous decay for non-interacting field theories. Later we show that this technique can be modified to obtain imaginary part of two-point correlator, which describes induced pair production.

This paper is organized as follows. In section \ref{spont} we consider cases of spontaneous particle creation and concentrate on $d=2$ dS space. We describe a method to deal with new type of divergences and derive formula for particle production rate in weak field limit. In section \ref{polarization} we calculate leading contribution to two-point correlator in  $d=2$ dS space. Finally in section \ref{dimentions} we generalize these cases to arbitrary $d$ and in section \ref{conclusion} we end up with conclusions.
 \section{Spontaneous pair production}
 \label{spont}

According to standard quantum field theory, the Euclidean one-loop effective action for a real massive scalar field $\phi$ with a self-interaction potential $U(f)$  can be written as

\begin{equation}
 \label{trlog}
\Gamma[\phi] = -\frac{1}{2} {\rm Tr\;\! ln} \left[ -\Box+m^2+U''(\phi)\right].
\end{equation}

We use the formula

\begin{equation}
 -{\rm Tr\;\! ln}\left( A \right) = \int\limits^\infty_0 \frac{dT}{T} {\rm Tr} \;e^{AT} ,
\end{equation}

remove $\phi(x)$ - independent term and perform functional trace in $x$-space. This gives us

\begin{equation}
 \Gamma[\phi] = \int\limits^\infty_0 \frac{dT}{T}  \int\limits d^D x \langle x | {\rm exp} \lbrace -T\left( -\Box + m^2 + U''(\phi) \right)\rbrace | x \rangle .
\end{equation}

The \textit{rhs} of this equation can be expressed in terms of path integral representation the same way we do it for solution of Schrodinger equation (look in \cite{csreview} for more details).

\begin{equation}
\Gamma [\phi] =
\int_0^{\infty}\frac{dT}{T}\, \!e^{-m^2T}\!
\int\limits_{x(1)=x(0)}
\; {\mathcal D}x \; {\rm exp}\left[-\int_0^1
\left(\frac{\dot x^2}{4T} +T U''(\phi(x)) \right)d\tau \right]\quad
\label{effective}
\end{equation}

There integration goes over all closed Euclidean spacetime paths $x^{\mu}(\tau)$ which are periodic in $\tau$ with period $1$. This effective action $\Gamma_{\rm Eucl} [\phi]$ is a functional of background classical field $\phi(x)$. We know that in similar case of scalar charged particle if classical background field  $A_{\mu}(x)$ corresponds to Minkowski electric field, the one-loop effective action has a nonperturbative imaginary part which is associated with pair production. Physical interpretation of this fact is that vacuum persistence amplitude is related with Minkowski effective action as

\begin{equation}
 \langle 0 | 0 \rangle = e^{i\Gamma_{Mink}},
\end{equation}

therefore imaginary part of $\Gamma_{Mink}$ can we identified with vacuum decay through pair production, such that

\begin{equation}
 P_{prod} = 1 - e^{-2{\rm Im}\Gamma_{Mink}} \approx 2{\rm Im}\Gamma_{Mink}.
\end{equation}

For example, in weak constant background electric field $E$ the expression for imaginary part is
\begin{equation}
\label{electric}
 {\rm Im}\,\Gamma_{\rm Mink}
\sim V_4^{\rm Mink}\,
\frac{e^2 E^2}{16\pi^3}
\,\exp\left[-\frac{m^2 \pi}{e E}\right],\quad
\end{equation}
and in the seminal work \cite{affleck} it was shown how to compute this leading contribution to imaginary part by using semiclassic approximation.

A similar effect also occurs in presence of non-flat background metric \cite{Morozov}\cite{KrotovPolyakov}. This problem is more often studied in terms of in-out vacua \cite{Bousso:2001mw} than in terms of worldline instantons \cite{Bastianelli:2002fv}. This inspires us to take advantage of worldline instanton method to study non-persistence of de Sitter space.

The main difference between Wick rotation of flat space and dS is that in the latter case topology gets changed from hyperboloid to a sphere:

\begin{equation}
\label{dS def}
\left\{
\begin{matrix}
ds^2 = dX_{d+1}^2  - \sum\limits_i dX_i^2\\
X_{d+1}^2 - \sum\limits_i X_i^2 = - R^2
\end{matrix}
\right.
\;\; \Rightarrow \;\;
\left\{
\begin{matrix}
ds^2 = - dX_{d+1}^2  - \sum\limits_i dX_i^2\\
X_{d+1}^2 + \sum\limits_i X_i^2 = + R^2
\end{matrix}
\right.
\;\; i=\overline{1,d}
\end{equation}

The same transformation as in (\ref{trlog})-(\ref{effective}) can be done for this space as well with some minor changes. Derivatives $\partial_\mu$ in (\ref{trlog}) should be replaced with covariant derivatives $\nabla_\mu$. It was shown in \cite{Grosche:1987de} that in (\ref{effective}) we get an additional integrand multiple $e^{T(\frac{d-1}{2r})^2}$. We can dispose of this constant term by using a shifted mass $\tilde m^2 = m^2 - \left( \frac{d-1}{2r}\right)^2 $. Let us start with simplest case of two-dimensional dS with zero background field. Then we can write our $\Gamma_{\rm Eucl}$ with nonzero imaginary part which appears due to background gravitational field following (\ref{effective})

\begin{equation}
\Gamma_{\rm Eucl} =
\int\limits_{0}^{\infty} \frac{dT}T e^{-\tilde m^2T }
\int\limits_{x_\mu(0)=x_\mu(1)} \; {\mathcal D} x_\mu(\tau)
{\rm exp}\left[-\int_0^1d\tau
\frac{\dot x^2}{4T} \right] .
\end{equation}

Using a method of steepest descent for path integral  corresponds to finding leading contribution for limit $mr \rightarrow \infty$. We find that action has several saddle points. They correspond to classical trajectories $l$ times winded on equator. The following step is to split path integral over closed trajectories into integral with Dirichlet boundary conditions and integration over boundary conditions. Probability of decay in homogenous space is proportional to it's volume, so we want to detach this factor from the very beginning. Path integral in

\begin{equation}
\label{EuclG}
\Gamma_{\rm Eucl}  = \int\limits_{0}^{\infty} \frac{dT}T e^{-\tilde m^2T }\int d^2 \tilde{x} \int\limits_{x_\mu(0)=x_\mu(1)} {\mathcal D}x_\mu(\tau) \; \delta^{(2)}(x_\mu(0)-\tilde{x})\; {\rm exp} \; \left[-\int_0^1d\tau \frac{\dot x^2}{4T} \right]
\end{equation}

doesn't depend on $\tilde{x}$, so we can replace it with $x = 0$. For computational purposes it is convenient to introduce standard angular coordinates

\begin{equation}
\label{spherical coords}
 x^{\mu} = \left(\begin{matrix} r\phi \\ r\theta \end{matrix}\right)\quad X_0 = r\sin(\theta)\cos(\phi) \quad X_1 = r\sin(\theta)\sin(\phi) \quad X_2 = r\cos(\theta).
\end{equation}

In this coordinate system the classical trajectories are

\begin{equation}
\label{ClassicTraj}
 X^{\mu}_{(l)}(\tau)= \left(\begin{matrix} 2\pi r \tau l \\ \frac\pi2 r \end{matrix}\right).
\end{equation}

As our boundary conditions are periodic, we can decompose $x^{\mu}(\tau)$ near these trajectories in Fourier series:

\begin{equation}
\label{fourierx}
 x^{\mu}(\tau) = X_{(l)}^{\mu} + \frac{\alpha_0}{\sqrt{T}} + \sqrt{\frac{2}{T}} \sum\limits_{n=1}^{\infty} \alpha^{\mu}_n \cos (2\pi\tau n) + \tilde\alpha^{\mu}_n \sin (2\pi\tau n).
\end{equation}

This normalization will prove to be somewhat convenient in future, but naturally physical answers don't depend on it. We can get rid of delta-functions via standard trick $\delta^{(2)}(r_\mu) = \displaystyle\int\frac{d^2p_\mu}{(2\pi)^2} e^{-ip_\mu x^\mu}$. Introducing
\begin{equation}
S_{1\delta} = \int_0^1d\tau \frac{\dot x^2}{4T} + ip_\mu x^\mu(0),
\end{equation}
answer for (\ref{EuclG}) in steepest descent approximation is

\begin{equation}
\label{descentGamma}
\Gamma_{\rm Eucl}  = -V \int\limits_{0}^{\infty} \frac{dT}T e^{-\tilde m^2T }  \sum\limits_{l} \left(\det \delta^2 S_{1\delta}\left(X_{(l)}\right)\right)^{-\frac12} e^{-S\left(X_{(l)}\right)},
\end{equation}
where $V$ denotes volume of observed system. A note should be made about measure of path integral. In curved spacetime measure
\begin{equation}
\label{measure}
 {\mathcal D} x(\tau) = N\prod\limits_{i} dx^{\mu}(\tau_i) \sqrt{g\left(x(\tau_i)\right)},
\end{equation}
 includes metrics determinant $\sqrt{g(x)}$, which equals $1$ on trajectory (\ref{ClassicTraj}). By $\delta$ we denote variation by all $\alpha^\mu_n$, $\tilde\alpha^\mu_n$ and $p^\mu$ as measure (\ref{measure}) is product of integration over fourier coefficients
 \begin{equation}
\label{measure2}
 {\mathcal D} x(\tau) = N\prod\limits_{n,\mu} d\alpha_n^\mu d\tilde\alpha_n^\mu .
\end{equation}
  Integral near trivial saddle point $X_{(0)}$ doesn't feel any effects of compact or curved space, so to define normalizing multiplier $N$ we can choose a default regularization scheme \cite{InstABC} for it

\begin{equation}
\label{regularization}
 \left(\det \delta^2 S_{1\delta}\left(X_{(0)}\right)\right)^{-\frac12} e^{-S\left(X_{(0)}\right)} = tr \; {\rm exp} \left(-\frac{p^2}2 T\right) = \frac{1}{2\pi T}.
\end{equation}

From practical point of view we will be interested only in two first terms in (\ref{descentGamma}) because classical action $S\left(X_{(l)}\right)$ is proportional to $l$ and, therefore, other terms will be exponentially suppressed and $l=0$ gives us the same result as in flat space and doesn't contribute to ${\rm Im}\Gamma$. We need a negative mode in $\delta^2 S_{1\delta}\left(X_{(0)}\right)$ (or any odd number of negative modes, to be precise). Expanding $S_{1\delta}$ near $X_{(0)}$

\begin{equation}
\label{quadraticl0}
 S_{1\delta} = 0 + \frac{1}{T^2} \sum\limits_{n=1}^{\infty} \pi^2 n^2 \left((\alpha^{\mu}_n)^2 + (\tilde\alpha^{\mu}_n)^2  \right) + i\frac{\sqrt{2}}{\sqrt{T}}\sum\limits_{n=1}^{\infty} \left( \alpha^\mu_n p_\mu \right) + ip_\mu\frac{\alpha^\mu_0}{\sqrt{T}}
\end{equation}

we see that $\delta^2 S_{1\delta}\left(X_{(0)}\right)$ has a positive-defined form and is also split into two independent pieces corresponding to two orthogonal directions $\mu = 1,2$. This means that determinant splits into product of independent parts corresponding to this directions
\begin{equation}
\det \delta^2 S_{1\delta}\left(X_{(0)}\right) = \det \delta_1\delta_1 S_{1\delta}\left(X_{(0)}\right) \det \delta_2 \delta_2 S_{1\delta}\left(X_{(0)}\right).
\end{equation}
 As we see from expansion near $X_{(1)}$

\begin{eqnarray}
\label{quadraticl1}
 S_{1\delta} = \frac{ \pi^2 r^2}{T} + \frac{1}{T^2} \sum\limits_{n=1}^{\infty} \pi^2 n^2 \left( (\alpha^{1}_n)^2 + (\tilde\alpha^{1}_n)^2 + ((\alpha^{2}_n)^2 + (\tilde\alpha^{2}_n)^2)(1-\frac1{n^2}) \right) - \frac{\pi^2 (\alpha_0^2)^2}{T^2} + \nonumber \\
 i\frac{\sqrt{2}}{\sqrt{T}}\sum\limits_{n=1}^{\infty} \left( \alpha^\mu_n p_\mu \right) + ip_\mu\frac{\alpha^\mu_0}{\sqrt{T}}
\end{eqnarray}

 latter statement about determinant separation remains valid. We also gain a negative eigenvalue. However, there is a zero eigenvalue which was not present in flat case - quadratic part is independent of $\tilde\alpha_1^2$. We deal with negative mode as in \cite{colemanCorrections}

\begin{equation}
\label{ColemanTransformation}
\int\limits_{-\infty}^{-\infty}dz e^{|a|z^2} \Rightarrow \frac i2 \int\limits_{-\infty}^{-\infty}dz e^{-|a|z^2}.
\end{equation}

Integral over $\alpha_1^2$ has a clear geometric interpretation. Even when we fix boundary conditions $x^\mu(0) = x^\mu(1) =  0$, there is still more than one classical trajectory - $SO(3)$ 2-sphere symmetry group leaves action intact. However, we can parameterize them by the point of intersection with transversal equator $x^1=\pm\frac{r\pi}{2}$. So seemingly infinite integral $d\alpha_1^2$ is integral over compact $S^1$ with conversion factor

\begin{equation}
\label{compactReplacement}
 d\alpha_1 = d\tilde{\phi}\;\! r\sqrt{\frac{T}{2}},
\end{equation}

which we derive from (\ref{fourierx}). Of course every classic solution intersects with transversal equator in two points - $\tilde{\phi}$ and $\tilde{\phi} +\pi$ but every path has two directions so we must integrate over all $\tilde{\phi}$.

With this remarks the following formula is pretty straightforward.
\begin{equation}
\label{compositionGammaEucl2}
{\rm Im}\;\Gamma_{\rm Eucl} = \int\limits_{0}^{\infty} \frac{dT}T e^{-\overline m^2 T - S_{1\delta}\left(X_{(1)}\right) } \left(\det \delta_1\delta_1 S_{1\delta}\left(X_{(1)}\right) \det ' \delta_2 \delta_2  S_{1\delta}\left(X_{(1)}\right)\right)^{-\frac12} \frac{1}{\sqrt{\pi}}d\tilde{\phi}\;\! r\sqrt{\frac{T}{2}},
\end{equation}
where $\det '$ denotes determinant for which zero modes have been thrown out and negative ones deal with as in (\ref{ColemanTransformation}) (see appendix \ref{appDets} for more details). After taking last integration over $T$ we arrive with
\begin{equation}
{\rm Im}\;\Gamma_{\rm Eucl} = \int\limits_{0}^{\infty} \frac{dT}T e^{-\tilde m^2 T - \frac{\pi^2r^2}{T} } \; \frac{Vr \sqrt {\pi}}{2T^{\frac 32 }} = \frac {V}{4\pi^2r^2} (1+2\pi \tilde mr) e^{-2\pi \tilde mr} \approx
\frac {4\tilde mV}{\pi r} e^{-2\pi \tilde mr}.
\end{equation}

Exponential part of this formula is consistent with \cite{KrotovPolyakov}.

 \section{Imaginary part of two-point scalar function in de Sitter}
\label{polarization}

The goal of this chapter is to find leading imaginary contribution to one-loop two-point amplitude in massive theory with potential
\begin{equation}
U(\phi)=\frac{\lambda}{3!} \phi^3.
\end{equation}
 It defines decay rate of scalar particles into two due to in-loop interaction with background gravitational field. We will start with two dimensions, but in chapter \ref{dimentions} we will discuss other cases. Two-point amplitude can be defined as \cite{Monin:2010zz},\cite{Satunin:2013an} a variational derivative of statistical sum $Z[\phi] = e^{\Gamma[\phi]}$
\begin{equation}
 \Pi(y,z) = \frac1Z \frac{\delta}{\delta \phi (y)} \frac{\delta}{\delta \phi (z)} Z = \frac{\delta}{\delta \phi (y)} \frac{\delta \Gamma}{\delta \phi (z)} + \frac{\delta \Gamma}{\delta \phi (y)}\frac{\delta \Gamma}{\delta \phi (z)}.
\end{equation}

Obviously second contribution doesn't depend on distance between points so we will discard it right away. With knowledge that

\begin{equation}
 \frac{\delta}{\delta \phi (y)}\int_0^1 \phi(x(\tau)) d\tau = \int_0^1 \delta^2 (x-y) d\tau
\end{equation}

it can be rewritten as
\begin{equation}
 \Pi (y,z) = \lambda^2\int\limits_{0}^{\infty} \frac{dT}T e^{-\tilde m^2T } \int  \frac{d^2 p d^2 q }{ (2\pi)^4 }T^2 \int\limits_0^1 d\tau' \int\limits_0^1 d\tau'' \int {\mathcal D}x(\tau) \;  e^{S_{2\delta}-ipy-iqz},
\end{equation}

 where in
 \begin{equation}
 S_{2\delta} =  \int_0^1d\tau \frac{\dot x^2}{4T} + ip_\mu x^\mu(\tau') + iq_\mu x^\mu(\tau'')
 \end{equation}

 we have gathered all terms with quadratic contributions. It also contains linear contribution $ip_\mu X^\mu_{(1)}(\tau') + iq_\mu X^\mu_{(1)}(\tau'')$.

The path integral in our approximation is still of a Gauss type, but with a linear shift

\begin{equation}
\label{scetchGauss}
\int d\vec\alpha\; exp\left( - \vec\alpha^T\delta^2 S \vec\alpha - i\vec J \vec \alpha \right) = \left(\det \delta^2 S\right)^{-\frac12} exp\left( - \frac14 \vec J\left(\delta^2 S\right)^{-1} \vec J^T  \right).
\end{equation}

This notation is rather sketchy but gives general idea - vector of collective coordinates $\vec\alpha$ contains everything we integrate out with infinite limits. In order to simplify our task we can choose particular points $y$ and $z$

\begin{equation}
\label{choiseyz}
 y^{\mu} = \left(\begin{matrix} r\phi_1  \\ \frac\pi2 r \end{matrix}\right) \quad z^{\mu} = \left(\begin{matrix}  r \phi_2 \\ \frac\pi2 r \end{matrix}\right),
\end{equation}

as we know that this correlator depends only on distance $r(\phi_2-\phi_1)$ between $y$ and $z$. The determinant for path integral

\begin{equation}
\det ' \delta^2 S_{2\delta}\left(X_{(1)}\right) = 4T^4\sin^2(2\pi\tau)\tau(1-\tau)
\end{equation}

does not depend on $\tau'$ and $\tau''$ but only on difference $\tau = \tau' - \tau''$. It is zero for $\tau=0,\frac12,1$ (for more details see appendix \ref{appDets}). It is worth noting that while for $\tau=\frac12$ we can regularize this determinant the same way we did in chapter \ref{spont}

\begin{equation}
\label{detexeption12}
\left(\det ' \delta^2 S_{2\delta}\left(X_{(1)}\right)_{\tau=\frac12}\right)^{-\frac12} \int d\alpha = \left(\frac{2}{\pi^2}T^6\tau(1-\tau)\right)^{-\frac12}\int d\tilde{\phi}\;\! r\sqrt{\frac{T}{2}},
\end{equation}
it has no negative eigenvalues and does not contribute to imaginary part of the correlator. The geometric reason for this is quite evident: $\tau=\frac12$ corresponds to all trajectories passing through two opposite points on a sphere. While there is a set of them with the same action, you can't deform it into a shorter  one.
Integrating over ${\mathcal D}x^1(\tau)$, we get a quadratic contribution to effective action

\begin{eqnarray}
 -\frac 14\vec J_1 \left(\delta_1^2S_{2\delta}\left(X_{(1)}\right)\right)^{-1} \vec J_1^T = \frac{-r^2\left((\phi_1-2\pi\tau')^2+(\phi_2-2\pi\tau'')^2\right) }{4T\tau(1-\tau)}.
\end{eqnarray}

Taking  integral over $d\tau'$ and $d\tau''$ using steepest descend method, the minimum of the exponential part is in
\begin{equation}
\tau'=\frac{\phi_1}{2\pi}\quad\quad \tau''=\frac{\phi_2}{2\pi},
\end{equation}
where this contribution to action vanishes. The resulting multiplier is $\frac{1}{1}4\pi T\tau(1-\tau)$. Integrals over $dT$ are
\begin{eqnarray}
 \int\limits_0^\infty dT e^{-m^2T-\frac{\pi^2 r^2}{T}}  \approx \frac{\pi r}{m} e^{-2\pi mr}\sqrt{\frac{1}{mr}} \quad\quad\quad
 \int\limits_0^\infty \frac{dT}{\sqrt T} e^{-m^2T-\frac{\pi^2 r^2}{T}} =  e^{-2\pi mr}\frac{\sqrt\pi}{m}
\end{eqnarray}
for $\tau\neq\frac12$ and $\tau=\frac12$ respectively.

This gives us an answer for leading contribution of two-point correlator

\begin{eqnarray}
\label{correlatorAnswer}
 {\rm Im}\Pi(y,z)= \frac{\lambda^2}{16\pi^2(\tilde mr)^{\frac32}} \frac{\sqrt{|y-z|(2\pi r-|y-z|)}}{r|\sin(\frac{|y-z|}{r})|} e^{-2\pi \tilde mr}.
\end{eqnarray}

The pole at $|y-z|=\pi r$ contradicts direct calculation with determinant (\ref{detexeption12}), which gives us a correction to the real part

\begin{eqnarray}
 \Delta{\rm Re}\Pi\left(y,y+\left(\begin{matrix} \pi r  \\ 0 \end{matrix}\right)\right)= \frac{\lambda^2}{16\tilde mr}e^{-2\pi \tilde mr} .
\end{eqnarray}

Near $|y-z|=\pi r$ one of the $\delta^2 S_{2\delta}\left(X_{(1)}\right) $ eigenvalues goes to zero, so Gaussian approximation stops being valid. But then for this region $ \pi  - \frac{|y-z|}{r}\ll 1$ we can smoothen the correlator replacing $ |\sin(\frac{|y-z|}{r})|^{-1} \rightarrow {\rm Re}\left[\left(|\sin(\frac{|y-z|}{r})|+i\frac1{\pi\sqrt{\tilde mr}}\right)^{-1}\right]$

\section{Arbitrary dimension}
\label{dimentions}

The goal of this section is to apply previously developed methods to $d$-dimensional De Sitter space (\ref{dS def}). Introducing coordinates similar to (\ref{spherical coords})

\begin{equation}
\label{spherical coords d}
 x^{\mu} = \left(\begin{matrix} r\phi \\ r\vec\theta \end{matrix}\right)\quad X_0 = r\left(\vec\theta\right)\cos(\phi) \quad X_1 = rA\left(\vec\theta\right)\sin(\phi) \quad X_i = r\cos(\theta_i),
\end{equation}

we can obtain similar classic trajectories

\begin{equation}
X^{\mu}_{(l)}(\tau)= \left(\begin{matrix} 2\pi r \tau l \\ \frac\pi2 r \vec 1 \end{matrix}\right),
\end{equation}

which generalize (\ref{ClassicTraj}). Classical action for these trajectories doesn't depend on $d$. If we decompose action near $X^{\mu}_{(1)}(\tau)$

\begin{eqnarray}
\label{quadraticl1d}
 S_{1\delta} = \frac{ \pi^2 r^2}{T} + \frac{1}{T^2} \sum\limits_{n=1}^{\infty} \pi^2 n^2 \left( (\alpha^{1}_n)^2 + (\tilde\alpha^{1}_n)^2\right) + \frac{1}{T^2} \sum\limits_{n=1,\nu>2}^{\infty} \pi^2 n^2\left(((\alpha^{\nu}_n)^2 + (\tilde\alpha^{\nu}_n)^2)(1-\frac1{n^2}) \right) \nonumber \\- \frac{\pi^2 (\alpha_0^2)^2}{T^2} + i\frac{\sqrt{2}}{\sqrt{T}}\sum\limits_{n=1}^{\infty} \left( \alpha^\mu_n p_\mu \right) + ip_\mu\frac{\alpha^\mu_0}{\sqrt{T}},
\end{eqnarray}

 we get a sum of $d$ independent parts, $d-1$ of which are identical. This means that if we take path integral (\ref{EuclG}) with steepest descent method (\ref{descentGamma}), then $\det \delta^2 S_{1\delta}\left(X_{(l)}\right)$ factorizes as

\begin{equation}
  \det' \delta^2 S_{1\delta}\left(X_{(l)}\right) = \det \delta^2_1 S_{1\delta}\left(X_{(l)}\right) \left(\det' \delta^2_2 S_{1\delta}\left(X_{(l)}\right)\right)^{d-1}.
\end{equation}

 From procedure (\ref{ColemanTransformation}) every determinant $\det' \delta^2_2 S_{1\delta}\left(X_{(l)}\right)$ gives us a $i$ multiplier, therefore for odd $d$ all contributions are real and there is no pair production. In terms of Bunch-Davies vacuum it was noted in \cite{Bousso:2001mw}. That's why all further formulas are for even $d$. This is enough to give an answer for  ${\rm Im}\;\Gamma_{\rm Eucl}$ in form analogous to (\ref{compositionGammaEucl2}).

\begin{equation}
\label{compositionGammaEucld}
{\rm Im}\;\Gamma_{\rm Eucl} = V\int\limits_{0}^{\infty} \frac{dT}T e^{-\overline m^2 T - S_{1\delta}\left(X_{(1)}\right) } \left(\det \delta_1^2 S_{1\delta}\left(X_{(1)}\right) \det ' \delta_2^2  S_{1\delta}\left(X_{(1)}\right)\right)^{-\frac12} \int d\Omega_{d-1}\;\! \left(\frac{Tr^2}{2\pi}\right)^{\frac{d-1}2}.
\end{equation}

Here $d\Omega_{d-1}$ is measure an a $(d-1)$-dimensional sphere and $V$ is volume of space. Our reasoning for replacing $d-1$ integrals of zero modes with integration over compact sphere is the same as in section \ref{spont}. There is a manifold of classical solutions which can be parameterized by the point of intersection with sphere  $x^1=\pm\frac{r\pi}{2}$. Every trajectory crosses this sphere twice, but trajectory has two directions.

\begin{equation}
  \Gamma_{d} = 2^{1-\frac {3d}2} \pi^{\frac d2 - \frac12}r^{d-1}\frac{d}{\left(\frac d2 \right)!} \int\limits_0^\infty \frac{\sqrt T dT}{T^{d+1}} e^{-\tilde m^2T - \frac{\pi^2 r^2}{T}} \approx \frac{d}{\left(\frac d2 \right)!} \frac{2^{1-d}}{(2\pi)^{\frac d2}} \frac{\tilde m^{d-1}}{r} e^{-2\pi \tilde m r}.
\end{equation}

Dimensional part of pre-exponential factor $\frac{\tilde m^{d-1}}{r}$ differs from naive expectation $\left(\frac{\tilde m}{r}\right)^{\frac d2}$, which follows from analogy with (\ref{electric}). The reason for that can be explained in a following way. Physical dimension is fixed by volume. In case of electric field mass is absent from pre-exponential factor. Charge and field always enter the formulas as a product, so (\ref{electric}) has no other options. In our case effective charge $m$ and effective field $r$ can enter in different combinations. We study limit $mr\rightarrow \infty$ and look for leading contribution. There is a contribution proportional to $\left(\frac{\tilde m}{r}\right)^{\frac d2}$, but it is suppressed.

Leading contribution to imaginary part of two-point correlator can be also acquired in a similar way. With choice of arguments similar to (\ref{choiseyz}) we can use formula (\ref{correlatorAnswer}) if we replace $2$-dimensional determinant with $d$-dimensional. Once again this contribution is imaginary only for even $d$

\begin{eqnarray}
\det ' \delta^2 S_{2\delta}\left(X_{(1)}\right) =  \det ' \delta_1\delta_1 S_{2\delta}\left(X_{(1)}\right) \left(\det ' \delta_2\delta_2 S_{2\delta}\left(X_{(1)}\right)\right)^{d-1} = \\
\left(\frac{2T^2}{\pi}\right)^d \pi^2\tau(1-\tau)(\sin^2(2\pi\tau))^{d-1}\nonumber.
\end{eqnarray}

Integrals over $dT$ now have form

\begin{eqnarray}
 \int\limits_0^\infty \frac{dT}{T^{d-2}} e^{-m^2T-\frac{\pi^2r^2}{T}} = 2\frac{m^{d-3}}{(\pi r)^{d-3}}K_{d-3}(2\pi mr) \approx \frac{m^{d-3}}{(\pi r)^{d-3}} e^{-2m\pi r}\sqrt{\frac{1}{mr}}.
\end{eqnarray}

This brings us to an answer

\begin{eqnarray}
\label{correlatorAnswerD}
 {\rm Im}\Pi(y,z)= \lambda^2 \frac{\tilde m^{d-\frac 72}\pi^{2-2d}}{r^{d+\frac12} 2^{\frac{5d}{2}-1}}\frac{\sqrt{|y-z|(2\pi r-|y-z|)}}{|\sin(\frac{|y-z|}{r})|^{d-1}} e^{-2\pi\tilde mr} .
\end{eqnarray}

 \section{Conclusion}
\label{conclusion}

 We have studied a process of spontaneous and induced particle creation is an external gravitational field. We work with scalar field theory in de Sitter space. For probability of spontaneous pair production a derivation of pre-exponential factor has been presented. Among the advantages of the offered method is the simplicity of generalization to other dimensions. It also provides a clear geometrical interpretation of vanishing particle production in weak field limit for odd dimensions.

 This method was generalized to calculate leading contribution to two-point correlator, which describes particle decay. Unlike to equivalent process for scalar theory  in external electric filed, in de Sitter space exponential suppression compared of spontaneous and induced  pair production is equal.

{\bf Acknowledgements.}  The author would like to acknowledge discussions with A. Gorsky, E. Akhmedov and P. Satunin. This work was done under the partial financial support of RFBR-12-02-00284, RFBR-14-02-31768 and PICS-12-02-91052.
\appendix
\renewcommand\thesection{\Alph{section}}
\section{Determinant calculation} \label{appDets}

In this section of appendix we will give precise details on how to calculate infinite determinants which appear in our text. As we have already mentioned we chose regularization scheme (\ref{regularization}). This allows us to formally replace an infinite divergent product with some finite quantity. However to do so we first need to find such an infinite product for this defining example. As you can see from expansions (\ref{quadraticl0}) and (\ref{quadraticl1})
\begin{equation}
\det \delta_1\delta_1 S_{1\delta}\left(X_{(0)}\right) = \det \delta_2\delta_2 S_{1\delta}\left(X_{(0)}\right) = \det \delta_1\delta_1 S_{1\delta}\left(X_{(1)}\right) = 2\pi T.
\end{equation}
Longitudinal determinant isn't affected by the change of saddle point. To work with $\det \delta_1\delta_1 S_{1\delta}\left(X_{(1)}\right)$ as a matrix it is convenient to choose the following ordering of variables: $p^1,\alpha^1_0,\alpha^1_1, \tilde\alpha^1_1,\dots,\alpha^1_n, \tilde\alpha^1_n,\dots$. With such choice it can be written as
\begin{equation}
\delta_1\delta_1 S_{1\delta}\left(X_{(1)}\right) =
\begin{bmatrix}
0&\frac{i}{2\sqrt{T}}&\frac{i}{\sqrt{2T}}&0&\frac{i}{\sqrt{2T}}&0&\dots\\
\frac{i}{2\sqrt{T}}&0&0&0&0&0&\dots \\
\frac{i}{\sqrt{2T}}&0&\frac{\pi^2}{T^2}&0&0&0&\dots \\
0&0&0&\frac{\pi^2}{T^2}&0&0&\dots \\
\frac{i}{\sqrt{2T}}&0&0&0&\frac{4\pi^2}{T^2}&0& \dots\\
\dots&\dots&\dots&\dots&\dots&\dots&\dots
\end{bmatrix}.
\end{equation}
Although it is not diagonal, the structure is convenient enough. It can be split into sub-blocks
\begin{equation}
\begin{bmatrix} \mathbf{A}_{k\times k} & \mathbf{B}_{k\times\infty} \\ \mathbf{B}_{\infty\times k}^T & \mathbf{D}_{\infty\times\infty} \end{bmatrix},
\end{equation}
where block indices describe it's size. Then if we choose $k=2$ block $\mathbf{D}$ becomes diagonal and invertible. This qualities of  $\mathbf{D}$ will be valid for other matrices from this appendix, but with other parameter $k$. As $\mathbf{D}$ is diagonal and invertible, we can use Gaussian elimination and set matrix  $\mathbf{B}$ to zero without changing the determinant. The disadvantage we get is that matrix $\mathbf{A}$ is replaced with another matrix $\mathbf{\tilde A}$. Then determinant factorizes

\begin{equation}
\label{goodmatixform}
\left| \begin{matrix} \mathbf{\tilde A}_{k\times k} & \mathbf{0}_{k\times\infty} \\ \mathbf{C}_{\infty\times k} & \mathbf{D}_{\infty\times\infty} \end{matrix} \right| = \det \mathbf{\tilde A} \det \mathbf{D}.
\end{equation}

 On the positive side in case of $\delta_1\delta_1 S_{1\delta}\left(X_{(1)}\right)$ the only different element of matrices $\mathbf{\tilde A}$ and $\mathbf{A}$ is $\mathbf{\tilde A}_{1,1}$ and it has no effect on determinant. Therefore

\begin{equation}
\label{regularizationProduct}
\det \delta_1\delta_1 S_{1\delta}\left(X_{(1)}\right) =N\frac{1}{4T}\left(\prod\limits_{n=1}^{\infty} \frac{\pi^2n^2}{T^2} \right)^2 = 2\pi T
\end{equation}

we got a definition for our regularization. In case of

\begin{equation}
\delta_2\delta_2 S_{1\delta}\left(X_{(1)}\right)=
\begin{bmatrix}
0&\frac{i}{2\sqrt{T}}&\frac{i}{\sqrt{2T}}&0&\frac{i}{\sqrt{2T}}&0&\dots\\
\frac{i}{2\sqrt{T}}&-\frac{\pi^2}{T^2}&0&0&\dots \\
\frac{i}{\sqrt{2T}}&0&0&0&0&0&\dots \\
0 &0&0&0&0&0&\dots \\
\frac{i}{\sqrt{2T}}&0&0&0&\frac{3\pi^2}{T^2}&0&\dots \\
0 &0&0&0&0&\frac{3\pi^2}{T^2}&\dots \\
\dots&\dots&\dots&\dots&\dots&\dots&\dots
\end{bmatrix}
\end{equation}
 everything is slightly less straightforward. Minimal matrix $\mathbf{A}$ from (\ref{goodmatixform}) is for $k=4$. It has a single zero eigenvalue, corresponding to integration over $\tilde\alpha_1^2$. How to deal with such problems was explained in (\ref{compactReplacement}), but that also means that when compared to (\ref{regularizationProduct}) we have one Gaussian integral less. This means an additional multiplier $\frac1{\sqrt{\pi}}$. Elements $\mathbf{A}_{1,1}$ and $\mathbf{\tilde A}_{1,1}$ also  differ, but it doesn't affect product of non-zero eigenvalues $-\frac{\pi^2}{2T^3}$ (determinant of upper-left minor $3\times3$). Keeping in mind procedure (\ref{ColemanTransformation}) with negative and zero eigenvalues we get

 \begin{equation}
\det' \delta_2\delta_2 S_{1\delta}\left(X_{(1)}\right) = 4 \left(\frac{2\pi^2}{4T^3}\right)\times\left(\prod\limits_{n=2}^{\infty} \frac{\pi^2(n^2-1)}{T^2} \right)^2 = \frac{4T^3}{\pi}.
\end{equation}

 In case of $S_{2\delta}$ determinants of $\mathbf{A}$ and $\mathbf{\tilde A}$ are going to be different. Due to second $\delta$-function we get a new integration variable $q_\mu$, therefore size of matrix $\mathbf{A}$ becomes $3\times3$ for $\delta_1\delta_1 S_{2\delta}\left(X_{(1)}\right)$ and $5\times5$ for $\delta_1\delta_1 S_{2\delta}\left(X_{(1)}\right)$. Gaussian elimination zeroes block $\mathbf{B}$ for matrix
\begin{equation}
\delta_1\delta_1 S_{2\delta}\left(X_{(1)}\right) =
\begin{bmatrix}
0&0&\frac{i}{2\sqrt{T}}&\frac{i\cos(2\pi\tau')}{\sqrt{2T}}&\frac{i\sin(2\pi\tau')}{\sqrt{2T}}&\dots \\
0&0&\frac{i}{2\sqrt{T}}&\frac{i\cos(2\pi\tau'')}{\sqrt{2T}}&\frac{i\sin(2\pi\tau'')}{\sqrt{2T}}&\dots\\
\frac{i}{2\sqrt{T}}&\frac{i}{2\sqrt{T}}&0&0&0&\dots \\
\frac{i\cos(2\pi\tau')}{\sqrt{2T}}&\frac{i\cos(2\pi\tau'')}{\sqrt{2T}}&0&\frac{\pi^2}{T^2}&0&\dots \\
\frac{i\sin(2\pi\tau')}{\sqrt{2T}}& \frac{i\sin(2\pi\tau'')}{\sqrt{2T}} &0&0&\frac{\pi^2}{T^2}&\dots \\
\dots&\dots&\dots&\dots&\dots&\dots
\end{bmatrix},
\end{equation}

but it changes upper left $2\times2$ block of $\mathbf{A}$, which transforms into

\begin{equation}
\frac{T}{2}
 \begin{bmatrix}
  F(0)&F(\tau)\\
  F(\tau)&F(0)
 \end{bmatrix},
\end{equation}
  where $\tau=\tau'-\tau''$ and $F(\tau)=\sum\limits_{n=1}^{\infty} \frac{\cos(2\pi n\tau)}{\pi^2 n^2} = \frac16 - \tau + \tau^2$. Determinant

\begin{equation}
  \det \delta_1\delta_1 S_{2\delta}\left(X_{(1)}\right) = \frac{1-6F(\tau)}{24}\times \left(\prod\limits_{n=1}^{\infty} \frac{\pi^2n^2}{T^2} \right)^2 = \tau(1-\tau) 2\pi T^2
\end{equation}

 is proportional to $T^2$ which stands in agreement with dimensional analysis. For matrix

\begin{equation}
 \delta_2\delta_2 S_{2\delta}\left(X_{(1)}\right) =
\begin{bmatrix}
 0&0&\frac{i}{2\sqrt{T}}&\frac{i\cos(2\pi\tau')}{\sqrt{2T}}&\frac{i\sin(2\pi\tau')}{\sqrt{2T}}&\frac{i\cos(4\pi\tau')}{\sqrt{2T}}&\frac{i\sin(4\pi\tau')}{\sqrt{2T}}&\dots \\
0&0&\frac{i}{2\sqrt{T}}&\frac{i\cos(2\pi\tau'')}{\sqrt{2T}}&\frac{i\sin(2\pi\tau'')}{\sqrt{2T}}&\frac{i\cos(4\pi\tau'')}{\sqrt{2T}}&\frac{i\sin(4\pi\tau'')}{\sqrt{2T}}&\dots\\
\frac{i}{2\sqrt{T}}&\frac{i}{2\sqrt{T}}&-\frac{\pi^2}{T^2}&0&0&\dots \\
\frac{i\cos(2\pi\tau')}{\sqrt{2T}}&\frac{i\cos(2\pi\tau'')}{\sqrt{2T}}&0&0&0&0&0&\dots \\
\frac{i\sin(2\pi\tau')}{\sqrt{2T}}& \frac{i\sin(2\pi\tau'')}{\sqrt{2T}} &0&0&0&0&0&\dots \\
\frac{i\cos(4\pi\tau')}{\sqrt{2T}}&\frac{i\cos(4\pi\tau'')}{\sqrt{2T}}&0&0&0&\frac{3\pi^2}{T^2}&0&\dots \\
\frac{i\sin(4\pi\tau')}{\sqrt{2T}}& \frac{i\sin(4\pi\tau'')}{\sqrt{2T}} &0&0&0&0&\frac{3\pi^2}{T^2}&\dots \\
\dots&\dots&\dots&\dots&\dots&\dots&\dots&\dots
\end{bmatrix}
\end{equation}

changing upper left  $2\times2$ doesn't change determinant of $\mathbf{\tilde A}$, so

\begin{equation}
 \det ' \delta_2\delta_2 S_{2\delta}\left(X_{(1)}\right) = 4 \frac{\pi^2\sin^2(2\pi\tau)}{4T^4}\times \left(\prod\limits_{n=2}^{\infty} \frac{\pi^2(n^2-1)}{T^2} \right)^2 = \frac{2T^2\sin^2(2\pi\tau)}{\pi}.
\end{equation}

Multiplier 4 comes from transformation (\ref{ColemanTransformation}).

\section{Contraction of vectors with inverse matrix}
\label{appinverseFold}
In order to calculate two-point correlator in chapter \ref{polarization} we will need an answer for contraction in (\ref{scetchGauss}). As we stated in appendix \ref{appDets}, our choice for order of variables gave this matrices a convenient structure. We can use formula for blockwise inversion

\begin{equation}
 \mathbf{\Lambda}^{-1} = \begin{bmatrix} \mathbf{A} & \mathbf{B} \\ \mathbf{B}^T & \mathbf{D} \end{bmatrix}^{-1} = \begin{bmatrix} (\mathbf{A}-\mathbf{BD}^{-1}\mathbf{B}^{T})^{-1} & -(\mathbf{A}-\mathbf{BD}^{-1}\mathbf{B}^T)^{-1}\mathbf{BD}^{-1} \\ -\mathbf{D}^{-1}\mathbf{B}^T(\mathbf{A}-\mathbf{BD}^{-1}\mathbf{B}^T)^{-1} & \quad \mathbf{D}^{-1}+\mathbf{D}^{-1}\mathbf{B}^T (\mathbf{A}-\mathbf{BD}^{-1}\mathbf{C})^{-1}\mathbf{BD}^{-1}\end{bmatrix}
\end{equation}

if matrices $\mathbf{D}$ and $(\mathbf{A}-\mathbf{BD}^{-1}\mathbf{B}^T)$ are invertible. With our choice of variables we need to find only one contraction $\vec J \left(\delta_1\delta_1 S_{2\delta}\left(X_{(1)}\right)\right)^{-1} \vec J^T$, where
\begin{equation}
J=(\phi_1-2\pi\tau',\phi_2-2\pi\tau'',0,\dots).
\end{equation}
The only inverse block of $\delta_1\delta_1 S_{2\delta}\left(X_{(1)}\right)$ that is required is
\begin{equation}
(\mathbf{A}-\mathbf{BD}^{-1}\mathbf{B}^T)^{-1} =
\begin{bmatrix}
\frac{1}{T\tau(1-\tau)}&\frac{-1}{T\tau(1-\tau)}&-i\sqrt{T}\\
\frac{-1}{T\tau(1-\tau)}&\frac{1}{T\tau(1-\tau)}&-i\sqrt{T}\\
-i\sqrt{T}&-i\sqrt{T}&T^2(\tau^2-\tau+\frac13)\\
\end{bmatrix}.
\end{equation}

It quickly gives us the answer

\begin{equation}
\vec J_1 \left(\delta_1^2S_{2\delta}\left(X_{(1)}\right)\right)^{-1} \vec J_1^T = \frac{r^2\left((\phi_1-2\pi\tau')^2+(\phi_2-2\pi\tau'')^2\right) }{T\tau(1-\tau)}.
\end{equation}


\begin{thebibliography}{99}
\bibitem{colemanClassic}
  S.~R.~Coleman,
  ``The Fate of the False Vacuum. 1. Semiclassical Theory,''
  Phys.\ Rev.\ D {\bf 15}, 2929 (1977)
  [Erratum-ibid.\ D {\bf 16}, 1248 (1977)].
\bibitem{colemanCorrections}
  C.~G.~Callan, Jr. and S.~R.~Coleman,
  ``The Fate of the False Vacuum. 2. First Quantum Corrections,''
  Phys.\ Rev.\ D {\bf 16}, 1762 (1977).
\bibitem{Kobzarev:1974cp}
  I.~Y.~.Kobzarev, L.~B.~Okun and M.~B.~Voloshin,
  ``Bubbles in Metastable Vacuum,''
  Sov.\ J.\ Nucl.\ Phys.\  {\bf 20}, 644 (1975)
  [Yad.\ Fiz.\  {\bf 20}, 1229 (1974)].
\bibitem{Selivanov:1985vt}
  K.~B.~Selivanov and M.~B.~Voloshin,
  ``Destruction Of False Vacuum By Massive Particles,''
  JETP Lett.\  {\bf 42}, 422 (1985).
\bibitem{Schwinger} J.\,Schwinger, Phys.\,Rev.\,{\bf 82}, 664 (1951).
\bibitem{affleck}  I.~K.~Affleck, O.~Alvarez and N.~S.~Manton,
  ``Pair Production at Strong Coupling in Weak External Fields,''
  Nucl.\ Phys.\ B {\bf 197}, 509 (1982).
\bibitem{Dunne:2006st}
  G.~V.~Dunne, Q.~-h.~Wang, H.~Gies and C.~Schubert,
  ``Worldline instantons. II. The Fluctuation prefactor,''
  Phys.\ Rev.\ D {\bf 73}, 065028 (2006)
  [hep-th/0602176].
\bibitem{Bulycheva:2011gs}
  K.~Bulycheva and S.~Guts,
  ``Photon-Photon Interaction and Schwinger Process,''
  Phys.\ Rev.\ D {\bf 88}, 053001 (2013)
  [arXiv:1107.1428 [hep-ph], arXiv:1107.1428 [hep-ph]].
\bibitem{Monin:2009aj}
  A.~Monin and M.~B.~Voloshin,
  ``Photon-stimulated production of electron-positron pairs in electric field,''
  Phys.\ Rev.\ D {\bf 81}, 025001 (2010)
  [arXiv:0910.4762 [hep-th]].
\bibitem{Gorsky:2005yq}
  A.~Gorsky and M.~B.~Voloshin,
  ``Particle decay in false vacuum,''
  Phys.\ Rev.\ D {\bf 73}, 025015 (2006)
  [hep-th/0511095].


\bibitem{Morozov}  A.~Mironov, A.~Morozov and T.~N.~Tomaras, JETP Lett.\  {\bf 94}, 795 (2012)  [Pisma Zh.\ Eksp.\ Teor.\ Fiz.\  {\bf 94}, 872 (2011)] [arXiv:1108.2821 [gr-qc]].


\bibitem{KrotovPolyakov}  D.~Krotov and A.~M.~Polyakov, Nucl.\ Phys.\ B {\bf 849}, 410 (2011)  [arXiv:1012.2107 [hep-th]].
\bibitem{Akhmedov:2013xka}
  E.~T.~Akhmedov, F.~K.~Popov and V.~M.~Slepukhin,
  ``Infrared dynamics of the massive $\phi^4$ theory on de Sitter space,''
  Phys.\ Rev.\ D {\bf 88}, 024021 (2013)
  [arXiv:1303.1068 [hep-th]].

\bibitem{csreview}
C.~Schubert,
  ``Perturbative quantum field theory in the string inspired formalism,''
  Phys.\ Rept.\  {\bf 355}, 73 (2001)
  [hep-th/0101036].

\bibitem{Bousso:2001mw}
  R.~Bousso, A.~Maloney and A.~Strominger,
  ``Conformal vacua and entropy in de Sitter space,''
  Phys.\ Rev.\ D {\bf 65}, 104039 (2002)
  [hep-th/0112218].
\bibitem{Bastianelli:2002fv}
  F.~Bastianelli and A.~Zirotti,
  Nucl.\ Phys.\ B {\bf 642}, 372 (2002)
  [hep-th/0205182].
\bibitem{Grosche:1987de}
  C.~Grosche and F.~Steiner,
  ``The Path Integral On The Pseudosphere,''
  Annals Phys.\  {\bf 182}, 120 (1988).
\bibitem{InstABC} A. Vainshtein, V. Zakharov, V. Novikov and M. Shifman, Sov. Phys. Usp. 25, 195, 1982.



\bibitem{Monin:2010zz}
  A.~Monin and M.~B.~Voloshin,
  ``Semiclassical calculation of an induced decay of false vacuum,''
  arXiv:1004.2015 [hep-th].

\bibitem{Satunin:2013an}
  P.~Satunin,
  Phys.\ Rev.\ D {\bf 87}, 105015 (2013)
  [arXiv:1301.5707 [hep-th]].
\end{thebibliography}
\end{document}